\documentclass[prb,twocolumn,superscriptaddress,showpacs,
amsmath,amssymb]{revtex4}

\usepackage{graphicx} 
\usepackage{dcolumn} 

\begin{document}

\title{Repulsive Casimir forces at quantum criticality}

\author{P. Jakubczyk}
\email{pjak@fuw.edu.pl}
\affiliation{Institute of Theoretical Physics, Faculty of Physics, University of Warsaw, 
 Pasteura 5, 02-093 Warsaw, Poland}
\author{M. Napi\'{o}rkowski}
\affiliation{Institute of Theoretical Physics, Faculty of Physics, University of Warsaw, 
 Pasteura 5, 02-093 Warsaw, Poland}
 \author{T. S{\c e}k}
\affiliation{Institute of Theoretical Physics, Faculty of Physics, University of Warsaw, 
 Pasteura 5, 02-093 Warsaw, Poland}

\date{\today}

\begin{abstract}
We study the Casimir effect in the vicinity of a quantum critical point.  As a prototypical system we analyze the  $d$-dimensional imperfect (mean-field) Bose gas enclosed in a hypercubic container of extension $L^{d-1}\times D$ and subject to periodic boundary conditions. The thermodynamic state is adjusted so that $L\gg\lambda\gg D\gg\l_{mic}$, where $\lambda\sim T^{-1/2}$ is the thermal de Broglie length, and $l_{mic}$ denotes microscopic lengthscales. Our exact calculation indicates that the Casimir force in the above specified regime is generically repulsive and decays either algebraically or exponentially, with a  non-universal amplitude. 
\end{abstract}

\maketitle

\section{Motivation }
Casimir-type interactions \cite{Casimir48, Fisher78, Krech94, Mostepanienko97, Kardar99, Brankov00, Bordag01, Gambassi09} are nowadays recognized in a multum of systems spanning from biological membranes to cosmology. The QED and condensed-matter contexts are those, where the theoretical predictions concerning the existence and properties of Casimir forces found firm experimental confirmation \cite{Garcia99, Garcia02, Fukuto05, Ganshin06, Hartlein08}. Herein we specify to the latter context, where a fluctuating medium is enclosed in a hypercubic box of spatial extension $L^{d-1}\times D$. We assume $L\gg D$ throughout the study. The spectrum of fluctuations (both thermal and quantum) of the medium is constrained by boundary conditions imposed by the confining walls, and, as a result, the free energy acquires a contribution depending on the separation $D$. It therefore becomes favorable to either increase, or decrease the distance $D$, resulting in an effective interaction between the boundary walls. The experimentally confirmed cases usually correspond to situations, where the Casimir force is attractive. A  well-known exception is the case of two materials characterized by different dielectric properties \cite{Dzyaloshinskii61} . On the theory side, exact results place severe restrictions on the possibility of obtaining Casimir repulsion in QED models \cite{Kenneth06, Rahi10}. Detours around these restriction invoke out-of equilibrium systems \cite{Bimonte11, Messina11, Kruger11}. The situation is more complex in the case of condensed-matter systems, where, at least theoretically, one may change the character of the force by varying boundary conditions. 

In the present study, we consider an exactly soluble model of interacting bosons at finite, but asymptotically low temperature, in a thermodynamic state corresponding to the vicinity of a quantum critical point. We show that in a specific limit the Casimir force is repulsive and decays as a power of the separation $D$ even for the periodic boundary conditions which generically yield Casimir attraction. This gives a hint on the possible regime of parameters, where a repulsive Casimir force might be detectable experimentally in a system involving Bose-Einstein condensation, and, potentially a wider class of quantum-critical systems.       

The essential ingredient of the critical Casimir effect is the interplay between two large lengthscales: $D$ and the bulk correlation length $\xi$. As long as $\xi\ll D$, the effective interaction between the walls decays exponentially $F\sim e^{-D/\xi}$ with $\xi$ setting the decay scale. If, however, the system is tuned sufficiently close to a (bulk) critical point, or is in a phase exhibiting soft excitations, one has $\xi\gg D$, and $F\sim D^{-d+1}$. The crossover between the above two regimes [$D/\xi\ll 1$ and $D/\xi\gg 1$] is governed by a scaling function, showing universal properties. 

The situation becomes more complex for $T\to 0$, where, in addition to $L$, $D$ and $\xi$, the thermal de Broglie length $\lambda=\frac{h}{\sqrt{2\pi m k_B T}}$ becomes macroscopic. We assume here that a phase transition may by tuned by a non-thermal control parameter 
(such as density, pressure, or chemical composition). 
Considering the Casimir forces in the low-$T$ limit one  identifies three regimes differentiated by the hierarchy of the macroscopic scales $L$, $D$, and $\lambda$. The standard thermal regime is recovered for $L\gg D\gg \lambda$. For the case $\lambda\gg L\gg D$, where one performs the $T\to 0$ limit before sending the system size to infinity, by virtue of the quantum-classical mapping, one expects the system properties to be similar to those of the thermal regime, albeit in elevated dimensionality. Finally, there is the possibility of the thermal length being squashed between the scales characterizing the system size, namely 
\begin{equation}
L\gg \lambda \gg D\gg l_{mic} \;. 
\label{regime}
\end{equation}       
Here $l_{mic}$ denotes any microscopic length present in the system.
To our knowledge, the limit defined by Eq.~(\ref{regime}) was not addressed so far, and this is not very simple to give a prediction for the asymptotics of the Casimir force relying solely on general arguments. Note that Eq.~(\ref{regime}) implies that the thermodynamic limit cannot be taken the usual way, keeping temperature fixed. Instead, while increasing $D$ temperature has to be reduced so that the condition $\lambda\gg D$ remains fulfilled. On the other hand, from a realistic (and experimental) point of view, the hierarchy of Eq.~(\ref{regime}) corresponds to a perfectly well defined regime. 

In what follows, we analyze the Casimir forces in the limit of Eq.~(\ref{regime}), employing a specific microscopic model of interacting bosons, the so-called imperfect Bose gas (IBG), which exhibits a phase transition to a Bose-Einstein-condensed phase for $d>2$ at any $T\geq 0$. The transition can be tuned by varying the chemical potential, which acts as the non-thermal control parameter.   The model is susceptible to an exact analytical treatment within the grand-canonical formalism.  

\section{Model} 
 We consider a system of spinless, interacting bosons at a fixed temperature $T$ and the chemical potential $\mu$. The system is enclosed in a hypercubic box of volume $V=L^{d-1}\times D$  and is governed by the Hamiltonian 
\begin{equation}
 \hat{H}=\sum_{\bf k} \frac{\hbar^2{\bf k}^2}{2m}\hat{n}_{\bf k}+\frac{a}{2V}\hat{N}^2 \;, 
\label{Hamiltionian}
\end{equation} 
where we use the standard notation. The repulsive interaction term $H_{mf}=\frac{a}{2V}\hat{N}^2$ ($a>0$) may be recovered 
from a 2-particle potential $v(r)$ in the Kac limit $\lim_{\gamma\to 0}\gamma^dv(\gamma r)$, i.e. for vanishing interaction strength and diverging range. After imposing periodic boundary conditions, the grand canonical partition function is cast in the convenient form \cite{Napiorkowski11}: 
\begin{equation}
\label{GC}
 \Xi (T,L,D,\mu) = -i e^{\frac{\beta\mu^2}{2a}V}\left(\frac{V}{2\pi a \beta}\right)^{1/2} \int_{\alpha\beta-i\infty}^{\alpha\beta+i\infty}dse^{-V\phi(s)}\;, 
\end{equation}
where 
\begin{eqnarray}
\label{phi_def}
\phi(s)=-\frac{s^2}{2a\beta}+\frac{s\mu}{a}+\frac{1}{V}\sum_{n_d}\ln(1-e^{s-\beta\epsilon_{k_d}})- \\ \nonumber
\sum_{n_d}\frac{1}{D\lambda^{d-1}}\,\,g_{\frac{d+1}{2}}(e^{s-\beta\epsilon_{k_d}})\;.
\end{eqnarray}
Here $k_d=\frac{2\pi n_d}{D}$, $n_d\in\mathbb{Z}$, $\beta\epsilon_{k_d}=\frac{\lambda^2}{D^2}\pi n_d^2$, $g_n(z)=\sum_{k=1}^{\infty}\frac{z^k}{k^n}$ are the Bose functions, and the contour parameter $\alpha$ is negative. The occurrence of the factor $V$ in the exponential in Eq.~(\ref{GC}) assures that the saddle-point approximation becomes exact for $V\to\infty$:
\begin{equation}
\lim_{V\to\infty}\frac{1}{V}\log \Xi (T,L,D,\mu) = \frac{\beta\mu^2}{2a}-\phi (\bar{s})\;. 
\label{free_en}
\end{equation}
It follows that for $L\to\infty$ 
the problem of evaluating the partition function becomes reduced to solving the stationary-point equation 
\begin{equation} 
\label{saddle-point_eq}
\phi'(\bar{s})=0
\end{equation} 
for $s\leq 0$.
More explicitly Eq.~(\ref{saddle-point_eq}) reads:
\begin{eqnarray}
\label{saddle-point_eq_2}
\frac{\mu}{a}-\frac{\bar{s}}{a\beta}=\frac{1}{\lambda^{d-1} D}g_{\frac{d-1}{2}}(e^{\bar{s}})+ \frac{2}{V}\sum_{n=1}^{\infty}\frac{e^{\bar{s}-\pi\frac{\lambda^2}{D^2}n^2}}{1-e^{\bar{s}-\pi\frac{\lambda^2}{D^2}n^2}} + \nonumber \\ 
\frac{1}{V}\frac{e^{\bar{s}}}{1-e^{\bar{s}}} + \frac{2}{\lambda^{d-1} D}\sum_{k=1}^{\infty}\frac{e^{k\bar{s}}}{k^{\frac{d-1}{2}}}\sum_{n=1}^{\infty}e^{-k\pi\frac{\lambda^2}{D^2}n^2} \;\;\;.
\end{eqnarray}

Bulk properties of the model defined by Eq.~(\ref{Hamiltionian}) were studied rigorously since 1980s \cite{Davies72, Berg84, Lewis86, Zagrebnov01}. The limit $T\to 0$ and the (bulk) quantum critical behavior were addressed in Ref.~\cite{Jakubczyk13}. 
For $d>2$ in the phase diagram spanned by $\mu$ and $T$ there is a line of  second-order phase transitions to the phase hosting the Bose-Einstein condensate. The critical line extends down to $T=0$, where it ends with a quantum critical point. The transition at $T>0$ falls into the universality class of the spherical model \cite{Berlin52}. The Casimir forces  corresponding to the above model were investigated in Refs.~\cite{Napiorkowski11, Napiorkowski13} in the thermal regime, where $D\gg \lambda$. Here we analyze the opposite case defined by the relation (\ref{regime}). 

In addition to $D$ and $\lambda$,  the model involves the lengthscale 
\begin{equation}
L_\mu = (a|\mu|^{-1})^{1/d}\;, 
\label{l_mu}
\end{equation}
which may be large or small compared to $D$ and $\lambda$. The microscopic scale is set by 
\begin{equation}
l =\left(\frac{2 \pi m a }{h^2}\right)^{1/(d-2)} \;, 
\label{l_mic}
\end{equation}
which is defined for $d\neq 2$. The quantity $l$ plays the role of the microscopic scale $l_{mic}$ occurring in Eq.~(\ref{regime}). Observe, that both the above lengthscales involve the interaction coupling $a$ and therefore are not present  for the perfect Bose gas.

In the thermal regime the excess surface grand-canonical free energy (per unit area) $\omega_s(T,D,\mu)$ is extracted by a subtraction of the bulk contribution $\omega_b(T,\mu)$ from the full grand-canonical free energy $\Omega(T,L,D,\mu)=-\beta^{-1}\ln\Xi(T,L,D,\mu)$. One obtains: 
\begin{equation}
\omega_s(T,D,\mu)=\lim_{L\to\infty}\left[\frac{\Omega(T,L,D,\mu)}{L^{d-1}}-D\omega_b(T,\mu)\right]\;.
\end{equation}
The Casimir force (per unit area) is then evaluated via 
\begin{equation}
F=-\frac{\partial \omega_s(T,D,\mu)}{\partial D}\;. 
\end{equation} 
In the regime considered in the present paper [Eq.~(\ref{regime})] one can take the alternative approach amounting to calculating the derivative  $\partial_D \lim_{L\to \infty}\Omega(T,L,D,\mu)/L^{d-1}$ and neglecting a constant ($D$-independent) term identified with the    bulk contribution.
The character of the solution to Eq.~(\ref{saddle-point_eq}) in the asymptotic regime specified by (\ref{regime}) crucially  depends on the dimensionality $d$. Here we primarily focus on the interval $d\in [2,3]$, where the term $\frac{1}{\lambda^{d-1} D}g_{\frac{d-1}{2}}(e^{\bar{s}})$ in Eq.~(\ref{saddle-point_eq_2}) is singular at $\bar{s}\to 0^-$. This guarantees that when $L\to\infty$ (keeping all the other lengthscales fixed), the solution $\bar{s}(T,D,\mu)$ remains separated from zero. In consequence, the terms $\sim 1/V$ in Eq.~(\ref{saddle-point_eq_2}) can be dropped. The last contribution to  Eq.~(\ref{saddle-point_eq_2}) is bounded from above by $\sim \frac{1}{\lambda^{d-1} D}e^{-\pi \lambda/D}$ and is also negligible as compared to the other terms. The solution of Eq.~(\ref{saddle-point_eq_2}) then boils down to analyzing the equation 
\begin{equation}
\frac{\mu}{a}-\frac{\bar{s}}{a\beta}=\frac{1}{\lambda^{d-1} D}g_{\frac{d-1}{2}}(e^{\bar{s}})\;
\label{s_equation}
\end{equation}
in the asymptotic regimes, where the correlation length $\xi\sim |\bar{s}|^{-1/2}$ (see below) is asymptotically large.

\section{Results for $d=3$} 
For the case of $d=3$ the term $\frac{1}{\lambda^{d-1} D}g_{\frac{d-1}{2}}(e^{\bar{s}})$ in Eq.~(\ref{s_equation}) displays a logarithmic singularity at $\bar{s}\to 0^-$; $g_{1}(e^{\bar{s}})\approx -\ln |\bar{s}|$ for $|\bar{s}|\ll 1$. Eq.~(\ref{s_equation}) finds asymptotic solution in the following form: 
\begin{eqnarray}
\label{s_solution}
\bar{s} \approx \left\{ 
\begin{array}{l l}
  \frac{l}{D}\ln \frac{l}{D} & \quad \mbox{in Regime I } \\
  -e^{-\frac{\lambda^2D}{L_{\mu}^3}} & \quad \mbox{in Regime II  } \\ 
	-\frac{\lambda^2 l}{L_{\mu}^3}-\frac{l}{D}g_1(e^{-\lambda^2 l/L_{\mu}^3}) & \quad \mbox{in Regime III } \;. \\ \end{array}  \right. 
\end{eqnarray} 
Regime I corresponds to the condition  $L_{\mu}\gg (D\lambda^2)^{1/3}$; Regime II to $L_{\mu}\ll (D\lambda^2)^{1/3},$  $\mu>0$; and, finally, Regime III is defined by $L_{\mu}\ll (D\lambda^2)^{1/3}\;, \mu<0$. 

The crucial observation is that the quantity $|\bar{s}|$ is related to the correlation length $\xi$ by the formula \cite{Napiorkowski11, Napiorkowski12}
\begin{equation}
\xi = \kappa \lambda |\bar{s}|^{-1/2}\;,
\end{equation}
where $\kappa$ is a numerical constant.

It follows that the singularity of $\xi$ occurring at the quantum critical point (in the limit $T,\mu,D^{-1}\to 0$) is effectively cut off by the system width $D$ in Regime I, by the thermodynamic fields $T,\mu$ in Regime III, and be a combination of the thermodynamic and geometric parameters in Regime II. This gives rise to the rich behavior predicted for the Casimir force (bee below). 

\begin{figure}
 \includegraphics[width=8.7cm]{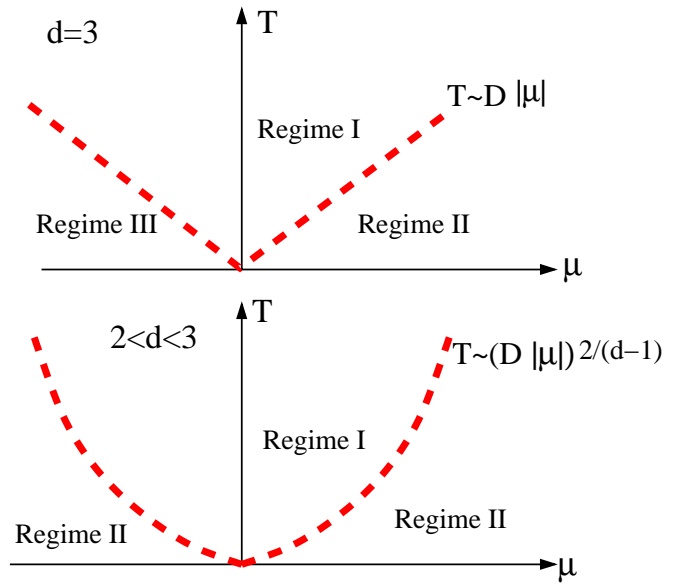} 
\caption{Ilustration of the asymptotic regimes occurring in $d=3$ (upper panel), and $2<d<3$ (lower panel). The asymptotic regimes are defined by Eq.~(\ref{regimes_definition}). The asymptotic behavior of the Casimir force is given by Eq.~(\ref{force_d_3}) in $d=3$ and Eq.~(\ref{force_d<3}) for $d\in ]2,3[$. In $d=2$ Regime I drops out, since the microscopic length $l$ of Eq.~(\ref{l_mic}) does not exist. }
\label{phase_diag}
\end{figure}

We now use Eq.~(\ref{free_en}) to compute the grand-canonical free energy $\Omega(T,L,D,\mu)$  and take the derivative  $\partial_D\lim_{L\to\infty} \Omega(T,L,D,\mu)$. Neglecting a constant, which is attributed to the bulk term in the free energy, we obtain the  following expressions for the Casimir force: 
\begin{eqnarray}
\label{force_solution}
\beta F(T,D,\mu) \approx \left\{ 
\begin{array}{l l}
  \frac{1}{2}\frac{l}{D^2\lambda^2}\ln^2(\frac{l}{D}) & \; \mbox{in Regime I} \\
  \frac{1}{L_{\mu}^3}e^{-\lambda^2D/L_\mu^3} & \; \mbox{in Regime II} \\ 
	\frac{1}{2}\frac{l}{D^2\lambda^2}g_1^2(e^{-\lambda^2/L_\mu ^3}) & \; \mbox{in Regime III}\;.   \\ \end{array}  \right. 
\label{force_d_3}
\end{eqnarray} 
The above result indicates that the effective force is always repulsive, and, except for Regime II, decays as a power of $D$. The logarithmic correction in Regime I and the exponential behavior in Regime II are specific to $d=3$. The obtained behavior is clearly different from that occurring in the thermal regime ($D\gg \lambda$), where the force is attractive and characterized by a universal amplitude wherever the interaction is long ranged (i.e. in the immediate vicinity of the transition or in the low-T phase). The present setup places the thermodynamic lengthscale $\lambda$ in between the macroscopic, geometric quantities $L$ and $D$, which has a far-reaching consequence for the properties of the Casimir interaction. In addition, the scale $L_\mu$ can be  adjusted at will, leading to the emergence of the three asymptotic regimes defined above. Also note that the Bose condensate, manifesting itself with the solution $\bar{s}=0$ (at $T$ finite) never appears in the analysis. This is because $D$ may not be made asymptotically large without sending temperature to zero (see Eq.~(\ref{regime}). In consequence, the condensate appears only in the strict limit of infinite $D$ and $T=0$. The results of this section are translated to the thermodynamic variables $\mu$, $T$ and summarized in Fig.~\ref{phase_diag}. 

\section{Results for $2<d<3$} 
It is interesting to follow the evolution of the system, and the Casimir forces in particular, when continuously reducing the dimensionality parameter $d$ from 3 towards the other physical value $d=2$. The case $d=3$ is special because the function $g_{\frac{d-1}{2}}(e^{\bar{s}})$ exhibits a logarithmic singularity at $\bar{s}\to 0^-$. For $d\in [2,3[$ we have:
\begin{equation}
g_{\frac{d-1}{2}}(e^{\bar{s}}) = \Gamma\left(\frac{3-d}{2}\right)|\bar{s}|^{\frac{d-3}{2}}+\zeta\left(\frac{d-1}{2}\right)+\dots\;.
\end{equation}
The asymptotic form of the saddle-point equation (\ref{s_equation}) admits the following solutions:
\begin{eqnarray}
\label{s_solution_2}
\bar{s} \approx \left\{ 
\begin{array}{l l}
  -C_1 l^{2\frac{d-2}{5-d}}\lambda^{-2\frac{d-3}{5-d}}D^{-\frac{2}{5-d}} &  \mbox{in Regime I } \\
  -C_2 L_{\mu}^{\frac{2d}{3-d}}\lambda^{-2\frac{d-1}{3-d}}D^{-\frac{2}{3-d}} &  \mbox{in Regime II  } \\ 
  - l^{d-2}\left(\lambda^2 L_\mu^{-d}+\lambda^{3-d}D^{-1}g\right) &  \mbox{in Regime III } , \\ \end{array}  \right. 
\end{eqnarray} 
where we introduced $C_1=(2\pi)^{-2/(5-d)}\left[\Gamma\left(\frac{3-d}{2}\right)\right]^{\frac{2}{5-d}}$, $C_2=\left[\Gamma\left(\frac{3-d}{2}\right)\right]^{\frac{2}{3-d}}$, and $g = g_{\frac{d-1}{2}}\left(e^{-\lambda^2 l^{d-2}L_\mu^{-d}}\right)$. The three emergent  asymptotic regimes are defined by the condition 
\begin{eqnarray}
\begin{array}{l l}
  L_\mu\gg \left(\lambda^{d-1}D\right)^{1/d} &  \mbox{Regime I} \\
  L_\mu\ll \left(\lambda^{d-1}D\right)^{1/d}\,\;\; \mu>0 &  \mbox{ Regime II} \\ 
  L_\mu\ll \left(\lambda^{d-1}D\right)^{1/d}\,\;\; \mu<0 &  \mbox{ Regime III} \;.   \\ \end{array}  
\label{regimes_definition}
\end{eqnarray} 
For $d=3$ this reduces to the previously obtained condition. 

From Eq.~(\ref{free_en}) we evaluate the free energy and extract the Casimir force by taking the $D$-derivative. The result reads: 
\begin{eqnarray}
\beta F(T,D,\mu) \approx \left\{ 
\begin{array}{l l}
  \frac{1}{2}C_1'^2 l^{\frac{(d-1)(d-2)}{5-d}}\lambda^{-2\frac{d-1}{5-d}}D^{-\frac{4}{5-d}} &  \mbox{ (I)} \\
  C_2 L_\mu ^{\frac{d(d-1)}{3-d}}\lambda^{-2\frac{d-1}{3-d}}D^{-\frac{2}{3-d}} &  \mbox{(II)} \\ 
  \frac{1}{2} l^{d-2} g^2_{\frac{d-1}{2}}\left(e^{\beta\mu}\right)\lambda^{-2(d-2)}D^{-2} & \mbox{(III)} \\ \end{array}  \right. 
\label{force_d<3}
\end{eqnarray} 
with $C_1'=[\Gamma(\frac{3-d}{2})]^{\frac{2}{5-d}}$.
The Casimir force is repulsive in all the three asymptotic regimes and decays with a power of $D$. Note a difference as compared to $d=3$, where the logarithms and exponents appeared as consequence of the form of the asymptotic behaviour 
of the Bose function $g_1$ at $\bar{s}\to 0^-$. As $d\to 3^-$, the power $\frac{2}{3-d}$ describing the decay of the force in Regime II [Eq.~(\ref{force_d<3})] diverges, which gives rise to the exponential behavior in $d=3$. The result is translated back to the thermodynamic 
variables $\mu$ and $T$ and depicted in Fig.~\ref{phase_diag}.

\section{Note on the case $d=2$}
We now comment on the Casimir force in $d=2$. This is a special case since the microscopic length $l_{mic}=l$ defined in Eq.~(\ref{l_mic}) does not exist. It makes no physical sense (nor mathematical) to consider the limit of the scales $D$ and $\lambda$ 
becoming macroscopic without specifying the microscopic length. The only choice possible in $d=2$ is to take $l_{mic}=L_\mu$ as given by Eq.~(\ref{l_mu}). The absence of the quantity $l$ in $d=2$ manifests itself in the non-existence of a solution to the 
saddle-point equation (\ref{saddle-point_eq}) in the parameter range corresponding to Regime I, where $L_\mu$ is large.

\section{Results for $d>3$} 
It is also interesting to examine the case $d>3$, where the system may host a thermodynamically stable Bose-Einstein condensate for $L\to\infty$, but finite $D$. For $d>3$ the function $g_{\frac{d-1}{2}}(e^{\bar{s}})$ is finite for $\bar{s}\to 0^-$. In consequence,  Eq.~(\ref{s_equation}) has no solution at sufficiently low $T$. This is because upon passing to the limit $L\to\infty$ in Eq.~ (\ref{saddle-point_eq_2}) $\bar{s}$ vanishes, and the term $\sim 1/V$ gives a finite contribution. It must therefore be included in the anaylsis by replacing Eq.~(\ref{s_equation}) with 
\begin{equation}
\frac{\mu}{a}-\frac{\bar{s}}{a\beta}=\frac{1}{\lambda^{d-1} D}g_{\frac{d-1}{2}}(e^{\bar{s}})+\frac{1}{V}\frac{e^{\bar{s}}}{1-e^{\bar{s}}}  \;. 
\label{new_s_eq}
\end{equation} 
The above equation is equivalent to the one arising in the bulk case for $D\gg \lambda$ [see Ref.~\cite{Napiorkowski13}] upon making the substitutions $\lambda^d\longrightarrow \lambda^{d-1}D$ and $g_{\frac{d}{2}}(e^{\bar{s}})\longrightarrow g_{\frac{d-1}{2}}(e^{\bar{s}})$. For $V\to\infty$ we find a finite, unique solution to Eq.~(\ref{new_s_eq}) provided $\lambda^{d-1}D\mu<a g_{\frac{d-1}{2}}(1)$. In the opposite case, the last term in Eq.~(\ref{new_s_eq}) gives a finite contribution equal to the condensate density \cite{Napiorkowski12}. This leads to the following result for the critical value of the chemical potential: 
\begin{equation}
\mu_c (T,D)= a \zeta\left(\frac{d-1}{2}\right)\frac{1}{\lambda^{d-1}D} \;, 
\end{equation} 
above which the Bose-Einstein condensate is present in the system. The transition between the phases may be induced by varying any of the parameters $\{\mu, T, D\}$ so that the geometric quantity $D$ may (for the presently relevant regime $D\ll \lambda$) serve as the tuning parameter on equal footing with the thermodynamic ones. One may also observe, that $\mu_c(T,D)$ may be related to the standard, thermodynamic critical value $\mu_c^{(d)}(T)$ of the chemical potential  \cite{Jakubczyk13} via 
\begin{equation}
\mu_c(T,D) =\mu_c^{(d)}(T)\frac{\lambda}{D}\frac{\zeta\left(\frac{d-1}{2}\right)}{\zeta\left(\frac{d}{2}\right)}\;. 
\label{muu_eq}
\end{equation}
Since $\lambda/D \gg 1$, it follows that  $\mu_c(T,D) > \mu_c^{(d)}(T)$. At $d\to 3^+$ we have  $\zeta\left(\frac{d-1}{2}\right)\to \infty$ in Eq.~(\ref{muu_eq}), so that $\mu_c(T,D)$ diverges and the condensate is ultimately suppressed in the regime $\lambda\gg D$.

For the determination of the Casimir force, we focus on the range of parameters, where the condensate is present ($\mu>\mu_c$), making the setup clearly distinct from that discussed for $d\leq 3$. The calculation leading to an expression for $\bar{s}$ is now analogous to the one performed in Ref.\cite{Napiorkowski13} (where one makes the replacements $\lambda^d\longrightarrow \lambda^{d-1}D$,  $g_{\frac{d}{2}}(e^{\bar{s}})\longrightarrow g_{\frac{d-1}{2}}(e^{\bar{s}})$ specified above). We obtain:
\begin{equation}
|\bar{s}(T,\mu,D)|\approx \frac{\lambda^{d-1}D}{V}\frac{1}{\zeta(\frac{d-1}{2})}\frac{\mu_c}{\mu-\mu_c}\;.
\end{equation}
This leads to the following expression for the Casimir force: 
\begin{equation}
\beta F(T,\mu,D) = \frac{4\pi}{\lambda^{d-3}D^3}g_{\frac{d-1}{2}}\left(e^{-\pi\lambda^2/D^2}\right)\;.
\end{equation}
which is again repulsive. 

\section{Summary} 
We have performed and exact study of Casimir forces occurring in the $d$-dimensional imperfect Bose gas (interacting bosons in the Kac limit) in the regime, where the de Broglie length is squashed in between the lengthscales $D$ and $L$ characterizing the system geometric size (i.e. for $D\ll \lambda \ll L$). We scanned the dependence of our results on the system dimensionality $d$. We obtain a behavior of the Casimir force completely different from that established before in the thermal regime (i.e. for $\lambda \ll D \ll L$) and also expected in the quantum regime $D \ll L \ll \lambda$ by virtue of the quantum-classical mapping. The computed Casimir force turns out to be repulsive and decay as a power of the distance $D$ in most of the cases (with log-corrections in $d=3$). 
 The peculiarity of our results may be traced back to the occurrence of an extra lengthscale ($\lambda$) which is considered as macroscopic, and which is absent in the standard condensed-matter setup. We emphasise that the present model perfectly fits into the established classification once we restrict the thermal regime (i.e. treat $\lambda$ as a microscopic scale). In this case it falls into the universality class of the $d$-dimensional spherical model.     
The interplay between $\lambda$, $D$, and the scale $L_\mu$ controlling the distance of the system from the (bulk) quantum critical point leads to the emergence of three regimes showing different asymptotic behavior of the Casimir force. An additional feature appears for $d>3$, where the system admits a $d-1$ dimensional (''surface'') condensate stable as a thermodynamic phase for any finite $D$. The transition to this phase may be tuned by varying $\mu$, $T$ as well as $D$.  It is interesting to speculate about the generality of our results and their dependence on the details of the microscopic model. Clearly, our results (for $\lambda \gg D$) do depend on the microscopic parameters. These however may be expressed via quantities of the dimensionality of length, which find natural analogues in other condensed-matter systems (in particular close to quantum criticality). We believe that  (when expressed via these length parameters) our finding should apply at least to other system belonging to the universality class of the spherical model \cite{Dantchev96, Dohm09, Diehl12, Dantchev14, Diehl14}. Another important question concerns the sensitivity of our results on the boundary conditions. Such a dependence is well known to occur in the thermal regime. We have checked that for von-Neuman boundary conditions the essential features of our results are unchanged.    

\acknowledgments
We thank Grzegorz \L ach, Anna Macio\l ek and Piotr Nowakowski for discussions, and Hans Diehl for a useful correspondence. We acknowledge funding from the National Science Centre via grant 2014/15/B/ST3/02212.

\end{document}